\documentclass[%
 aip,
 sd,%
 amsmath,amssymb,
 preprint,%
]{revtex4-1}

\usepackage{graphicx}
\usepackage{dcolumn}
\usepackage{bm}
\usepackage[colorlinks=true,linkcolor=blue,citecolor=blue]{hyper ref}
\usepackage{natbib}

\begin{document}

\title {Mechanical and optical nanodevices in single-crystal quartz}

\author{Young-Ik Sohn}
\email{sohn@seas.harvard.edu.}
\affiliation{John A. Paulson School of Engineering and Applied Sciences, Harvard University, 29 Oxford Street, Cambridge, Massachusetts 02138, United States}
\author{Rachel Miller}
\affiliation{John A. Paulson School of Engineering and Applied Sciences, Harvard University, 29 Oxford Street, Cambridge, Massachusetts 02138, United States}
\affiliation{Department of NanoEngineering, University of California San Diego, La Jolla, CA 92093, USA}
\author{Vivek Venkataraman}
\affiliation{John A. Paulson School of Engineering and Applied Sciences, Harvard University, 29 Oxford Street, Cambridge, Massachusetts 02138, United States}
\affiliation{Department of Electrical Engineering, Indian Institute of Technology Delhi, New Delhi, India}
\author{Marko Lon\v{c}ar}
\affiliation{John A. Paulson School of Engineering and Applied Sciences, Harvard University, 29 Oxford Street, Cambridge, Massachusetts 02138, United States}


\begin{abstract}
Single-crystal $\alpha$-quartz, one of the most widely used piezoelectric materials, has enabled a wide range of timing applications. Owing to the fact that integrated thin-film based quartz platform is not available, most of these applications rely on macroscopic, bulk crystal-based devices. Here we show that the Faraday cage angled-etching technique can be used to realize nanoscale electromechanical and photonic devices in quartz. Using this approach, we demonstrate quartz nanomechanical cantilevers and ring resonators featuring Qs  of 4,900 and 8,900, respectively.
\end{abstract}

\pacs{77.84.-s, 42.70.Ce, 77.65.Fs}
\keywords{quartz, NEMS, nanophotonics}
\maketitle

Silicon dioxide (SiO$_2$), the most abundant mineral found in the earth's crust, has eleven crystalline polymorphs determined by the temperature and pressure of the environment during the time of crystallization. Two of these are referred to as quartz: $\alpha-$quartz is stable below 573 $^\circ$C, and $\beta-$quartz is stable above this temperature. $\alpha-$quartz does not have a centro-symmetric crystal structure, \cite{newnham2004properties} which is the cause for its piezoelectric response that allows the coupling between electrical and mechanical degrees of freedom. Furthermore, owing to quartz's crystalline anisotropy, dozens of substrates in different cut planes can be realized. The plane of crystal cut determines the characteristics of quartz devices such as resonant frequency, temperature coefficient of frequency (TCF), stability and many others.\cite{Vittoz:2010uf} For example, AT-cut quartz is of great interest for temperature-insensitive crystal oscillators, whereas Z-cut is a common choice for tuning forks in watches.\cite{Friedt:2007bt} In the rest of this manuscript, we refer to $\alpha-$quartz as `quartz,' for simplicity.

Quartz is also a promising material for nanophotonic devices and systems. High purity silica glass, with its low loss and large transparency window, was the key element that enabled low-loss optical fiber technology and long-distance communications. Silica is also a popular material for nonlinear optics and nanophotonics.\cite{Agrawal:2013tm} Compared to amorphous silica, quartz has lower optical loss while having the advantage of fast tuning of refractive index via electro-optic effect.\cite{Li:2006bq} This combination makes the quartz a unique material for achieving ultra-high optical quality factor and electrical tuning simultaneously for the whispering gallery mode resonators.\cite{Ilchenko:2008wi}

In order to take a full advantages of quartz's remarkable material properties, it is important to enhance interaction between the material and electric field (optical or DC) which can be accomplished using nanoscale devices. However, high quality thin quartz films on foreign substrates are not available.\cite{Brinker:2013da} Yet, this is important for both electromechanical devices, where a sacrificial substrate is needed, and optical (and optomechanical) devices, where the substrate plays the role of cladding. Notable developments of the thin film platform for quartz include epitaxial growth on silicon substrates\cite{CarreteroGenevrier:2013tv} and wafer-to-wafer bonding of quartz and silicon.\cite{Imbert:2011ui} However, both methods have their own challenges. In order to realize devices in quartz, traditionally, wet etching of quartz has been most widely used. However, its anisotropic nature makes miniaturization difficult and causes unwanted features.\cite{Kamijo:ef} Deep reactive ion etching has been proposed as an alternative to etch along any crystal axis.\cite{Chapellier:hb} In this work, we demonstrate functional nanomechanical and photonic devices in bulk quartz crystals using modified dry etching approach.

Our approach (Figure 1a) is based on the Faraday cage angled-etching\cite{Latawiec:2016ca} that we have used to realize devices in bulk single-crystal diamond.\cite{Burek:2012bk} Specifically, standard bi-layer lift-off process was used to define an etch mask for cantilevers, while mask pattern was transferred from e-beam resist to sputtered metal film for microring resonators. The latter results in smoother mask and minimizes scattering loss of optical resonators, which is difficult to achieve with lift-off process. Etch parameters and the type of reactive-ion etching tools used are described in our previous work.\cite{Latawiec:2016ca} This approach can make fairly complex structures in quartz, including double-ended tuning fork (Fig. 1(b) and (c)) and microring resonators (Fig. 4). 

\begin{figure}[h]
        \centering
        \includegraphics[width=\columnwidth]{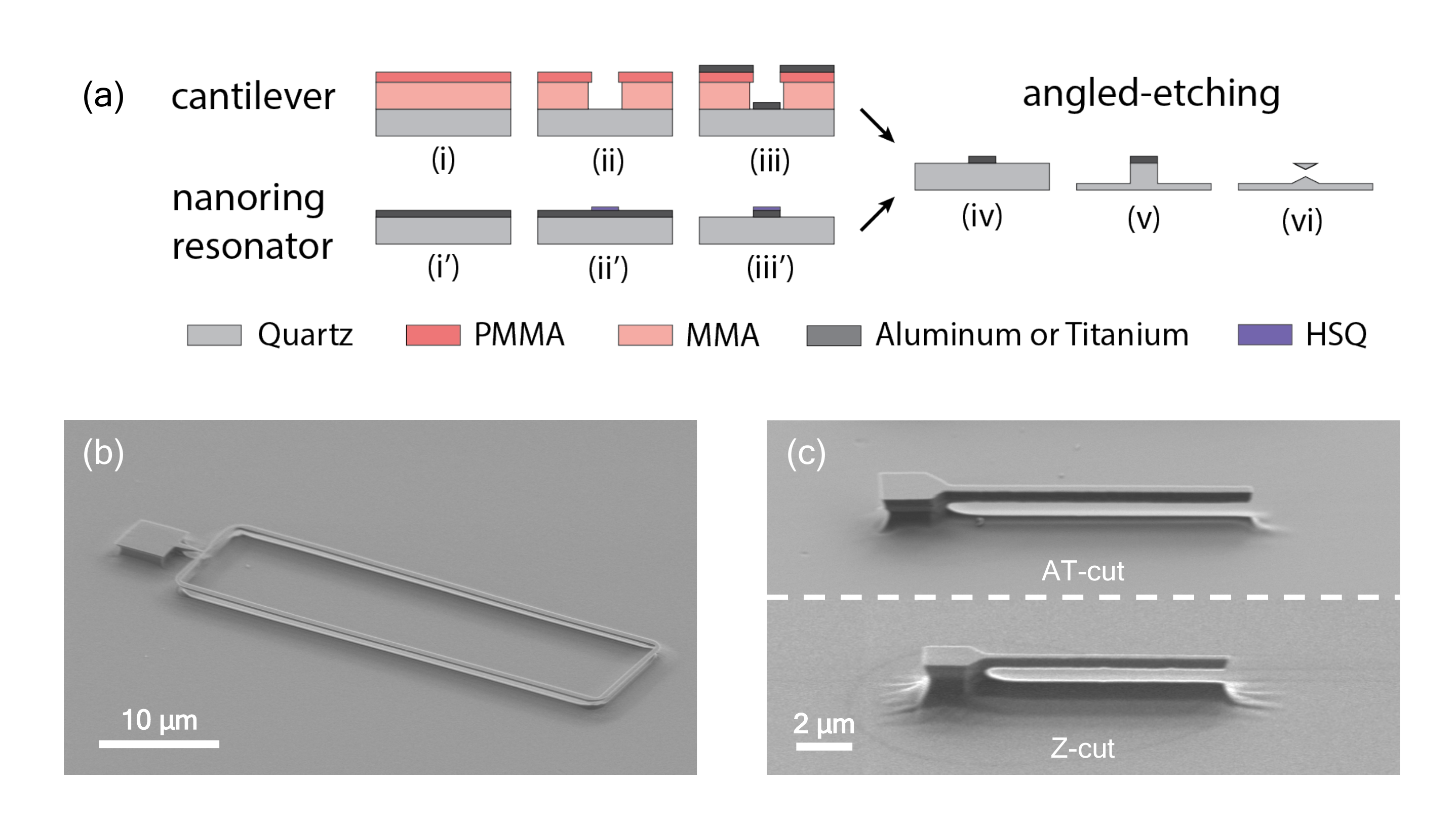}
        \caption{(a) Schematic illustration shows the fabrication process of cantilever and microring resonators. To define cantilevers, aluminum mask is deposited using standard bi-layer lift-off process. For ring resonators, titanium film was first sputtered on quartz, and then e-beam lithography and metal etching were used to define the mask. Next, angled-etching step is performed to fabricate (b) a double-ended tuning fork and (c) cantilevers, made in quartz with two different crystal cuts.}
        \label{FigureOne}
\end{figure}

 To show that our approach is applicable to different crystal cuts, cantilevers with similar dimensions have been fabricated by the same procedure in both AT-cut and Z-cut single-quartz crystals as shown in Fig. 1(c). We expect that our approach can be applied to all existing crystal cuts of quartz, as discussed previously in the literature.\cite{Vittoz:2010uf,Chapellier:hb}

\begin{figure}[h]
        \centering
        \includegraphics[width=\columnwidth]{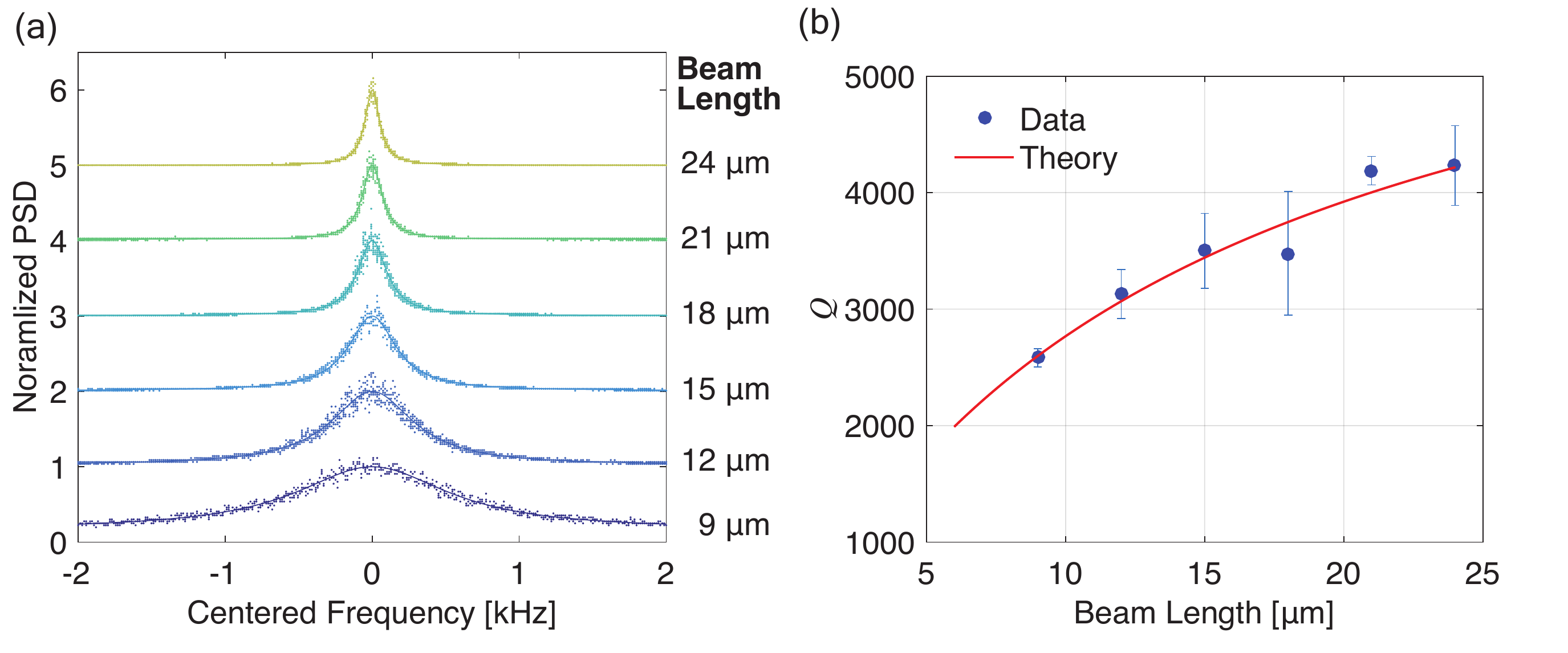}
        \caption{(a) Normalized power spectral density (PSD) of thermal fluctuations of cantilevers at different lengths. Longer cantilevers give better mechanical quality factors because of the smaller clamping loss. Dots are raw data and the solid lines are from the fit of Lorentzian functions. (b) A few dozen cantilevers are measured and their statistics of mechanical quality factors are fit to the model described in the main text. Vertical error bars represent standard deviations of measured cantilevers with the same length.}
        \label{FigureTwo}
\end{figure}

We fabricated twenty eight cantilevers, in the same Z-cut quartz substrate, having the same cross-section and lengths in the range of 9 - 24 $\mu$m. Neutral axis of all cantilevers is along $\langle 1\,1\,-2\,0 \rangle$ direction of the quartz crystal. We measured thermal fluctuations of cantilevers in its fundamental flexural mode by optical interferometric displacement detection.\cite{Karabacak:2005hq} Fitting Lorentzian function to thermal fluctuations, we can estimate the mechanical quality factor of the resonances. Fig. 2(a) shows thermal fluctuations of cantilevers with different lengths in normalized power spectral density, with shifted resonance frequencies. We note the trend of increasing mechanical quality factor with increasing length. To understand this trend and the mechanism that limits quality factors, we have taken statistics of all cantilevers with different lengths, where we have at least three cantilevers for each length. We plotted the average mechanical quality factors as a function of lengths in Fig. 2(b), with vertical bars indicating standard deviation of measured devices. Using the theory of mechanical loss of cantilevers, we can fit our data to the following model.\cite{Imboden:2014vf} 

\begin{equation}
Q^{-1}_{\text{total}} =  Q^{-1}_{\text{clamping}}  + Q^{-1}_{\text{other}} = K L^{-1} + Q^{-1}_{\text{other}}
\label{loss modeling}
\end{equation}
where $Q_{\text{total}}$ is the total mechanical quality factor we measure. The first and the second term on the right hand side account for the clamping loss and the rest of the loss mechanism, respectively. $L$ is the length of a cantilever and $K$ is a fitting parameter that is a constant. Clamping loss is well studied theoretically and experimentally. When the width and thickness of a cantilever are fixed, it is known that the loss rate is proportional to $L^{-x}$ where the value of exponent $x$ depends on the physical shape of clamping.\cite{Imboden:2014vf} By inspecting the shape of the clamp shown in Fig. 1(c), we assume that the clamp behaves similarly to that of out-of-plane motion with a large undercut,\cite{Imboden:2014vf} and therefore the exponent in the equation (\ref{loss modeling}) is assumed to be $-1$. The loss mechanism other than clamping loss is assumed to be independent of beam length. By fitting the equation (\ref{loss modeling}) to the data, we get an estimation of $Q_{\text{other}}=6,750\pm1,950$ (uncertainty is for 95\% confidence interval), and this length-independent loss is typically caused by the surface loss.\cite{Imboden:2014vf} Therefore to increase the mechanical quality factor, surface loss needs to be mitigated by, for example, surface treatment or device geometry with a low surface to volume ratio.

We further measured the temperature dependence of the resonance frequency of 12 $\mu$m long cantilever. Conventionally, quartz tuning forks can be used to make millimeter scale thermometers by using its thermal expansion and the temperature dependence of stiffness tensor. TCF is typically expanded in polynomial series up to the third order.\cite{2007PhST..129..316J}

\begin{equation}
f(T) = f(T_0) \left(1+\alpha(T-T_0)+\beta(T-T_0)^2+\gamma(T-T_0)^3\right)
\label{TCF}
\end{equation}
where $f$ is the resonance frequency and $T_0$ is the operating temperature. $\alpha$, $\beta$ and $\gamma$ are temperature coefficients of the first, second and third order, respectively. With a careful choice of crystal cut and the the direction of neutral axes of the tuning fork's tines, it is possible to make a temperature sensor of a good linear response (large $\alpha$ and small $\beta, \gamma$) over a wide frequency range.\cite{Ueda:1986kt}

Noting a flexural mode of a cantilever is fundamentally similar to that of tuning forks, we measured the temperature sensitivity of its resonance frequency. By increasing the temperature with steps of 5 $^\circ$C from 25 $^\circ$C to 100 $^\circ$C using a resistive heater and a closed-loop temperature controller (Thorlabs HT10K and TC200), we measured the response shown in Fig. 3(a). Fitting a linear curve gives the first order TCF of $-25.9 $ ppm/$^\circ$C. From the data, we calculated the deviation from the linearity as the difference between the fit model and measured temperatures. Fig. 3(b) shows that the deviations fall within $\pm 0.5^\circ$C.

\begin{figure}[h]
        \centering
        \includegraphics[width=\columnwidth]{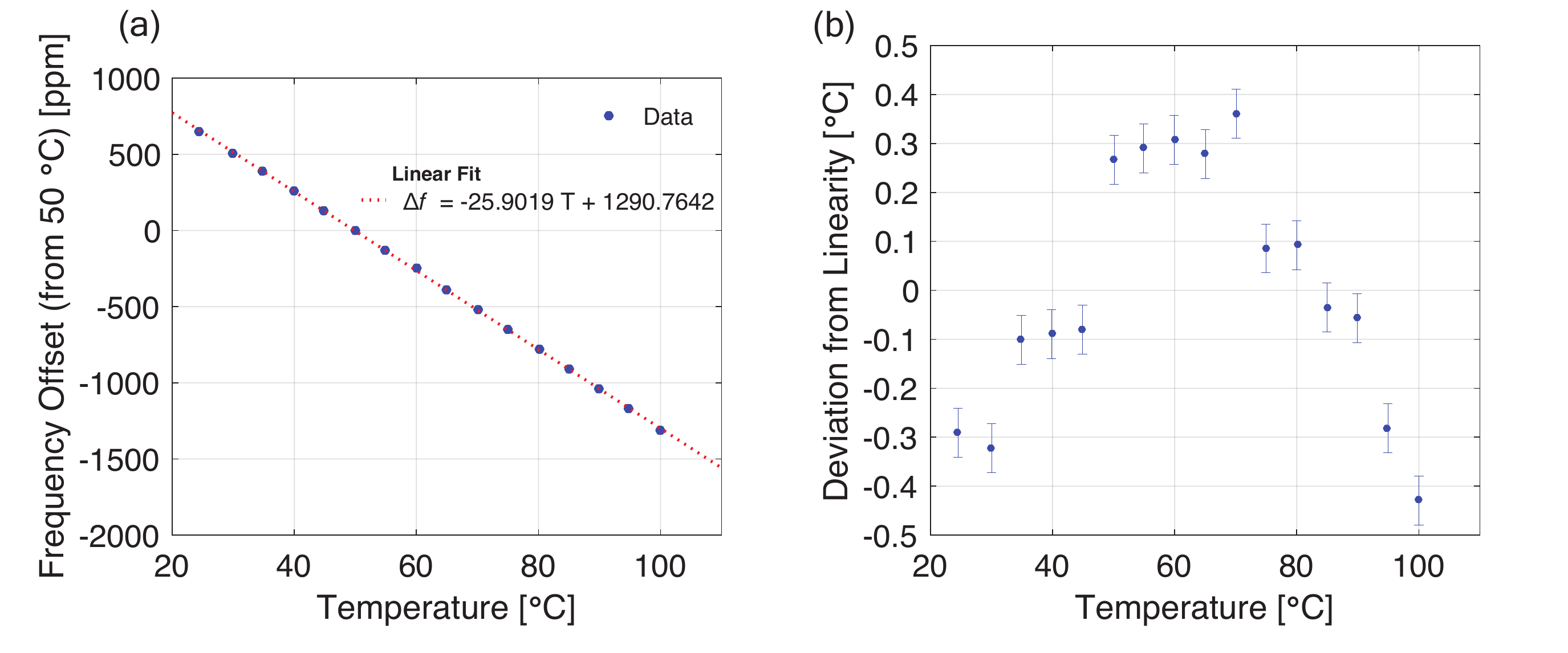}
        \caption{(a) Resonance frequency of the 12 $\mu$m long cantilever as a function of temperature. Linear fit gives the value of the first order TCF. Standard errors of frequency and temperature for each measurement are smaller than the size of dots, and hence they are not shown. (b) Deviation from the linearity in (a) are plotted to show the expected accuracy of temperature reading when the device is used as a thermometer. Vertical bars represent reading errors from digital readout of the temperature sensor.}
        \label{FigureThree}
\end{figure}

Nanophotonic devices can also be made using angled-etching, where air surrounding the structure is used as a cladding (Fig. 4(a)). Slight widening of the width in the straight part of the resonator provides structural support.\cite{Burek:2014bj} Width of 1.5 $\mu$m was used to target the operating wavelength range in telecom. Enlarged SEM images in Fig. 4(a) reveal the visible surface roughness. We measured its optical quality factor using a tapered fiber setup whose image is shown in Fig. 4(b). Fig. 4(c) shows the transmission of the fiber that has resonances as a series of dips which originate from the evanescent coupling between the resonator and the fiber. We extract quality factors of each resonance from their widths by fitting Lorentzian functions. All the resonances have loaded quality factors on the order of thousands, and the highest of them is $Q_{\text{total}}$ = 8,900 as shown in the inset of Fig. 4(c). From the fitting, we estimate the intrinsic quality factor of approximately $Q_{\text{intrinsic}}$ = 13,000.\cite{Spillane:2003jv} Quality factor at this level is lower than those of similar devices made on diamond.\cite{2017APLP....2e1301A,Latawiec:2016et,Burek:2014bj} Judging from images in Fig. 4(a), we believe that the limiting mechanism of optical quality factors is the surface roughness. By improving the fabrication process, (i.e. using a better etching mask material or dry etch recipe) we expect to increase optical quality factors.

\begin{figure}[ht]
        \centering
        \includegraphics[width=\columnwidth]{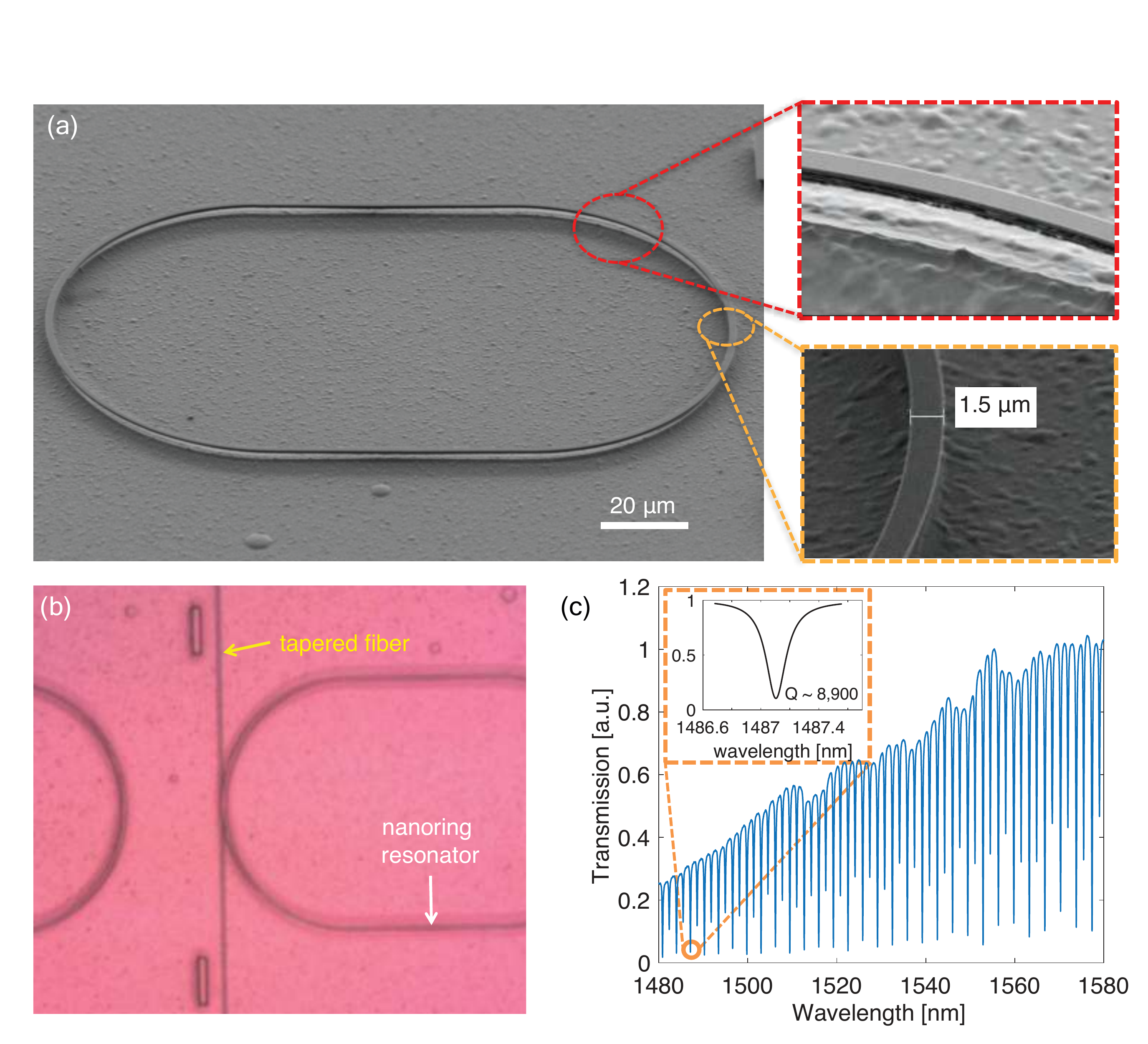}
        \caption{(a) SEM image of microring resonators whose operation wavelength range is telecom. (b) Microscope image of fiber coupling setup where the light couples from the single-mode tapered fiber to a microring resonator. (c) Transmission measurement taken from the setup pictured in (b). Dips correspond to resonance conditions and quality factors can be extracted from the widths of each dip. Loaded quality factors are on the order of thousands and the largest ($Q\sim 8,900$) is shown in the inset.}
        \label{FigureFour}
\end{figure}

In summary, we have applied the Faraday cage angled-etching technique to single-crystal quartz and fabricated suspended structures with substrates of different crystal cuts. First, we made cantilevers with nanoscale width and high aspect ratio. By analyzing the trend of mechanical quality factor as a function of cantilevers' lengths, we estimate the contribution of the surface loss to the quality factor to be on the order of thousands. With one of these cantilevers, we further measured the linear dependence of the resonance frequency on temperature. Finally, optical microring resonators for telecom wavelength were fabricated and their quality factors were measured. From the fitting, we estimated the intrinsic quality factor about ten thousands, where we presume the loss is limited by a rough surface.

In the future, to make devices of practical use, we propose several approaches in the following. First, the current etch recipe described in the reference\cite{Latawiec:2016ca} can be improved to make better devices. Surface quality of the fabrication in this work seems to be limiting the quality factors for both mechanical and optical devices. Therefore, a better fabrication process can potentially lead to higher quality factors for both types. Second, well-studied characteristics of different quartz crystal cuts can be exploited in combination with our approach to make various functional devices at the nanoscale, in a similar fashion to traditional engineering. For example, it is possible to engineer TCF by choosing different crystal cuts. Third, piezoelectric and electro-optic devices can be made by placing electrodes nearby. We have previously demonstrated the electrode patterning near diamond nanodevices made from angled-etching.\cite{Sohn:2015cv,Sohn:2017uw} Combining those electrodes with mechanical and optical elements in this work, we expect to fabricate more advanced piezoelectric or electro-optic nanodevices. \\

This work was supported by the STC Center for Integrated Quantum Materials, NSF Grant No. DMR-1231319. Samples were fabricated at the Center for Nanoscale Systems (CNS), a member of the National Nanotechnology Infrastructure Network (NNIN), which is supported by the National Science Foundation under NSF award no. ECS-0335765. CNS is part of Harvard University.

\bibliography{myrefs}

\end{document}